\documentclass{ws-procs9x6}

\begin{document}
\title{Spin Polarization of Hyperons 
in the Hadron-Hadron Inclusive Collisions
%\\
% - Mechanisms and Dynamics -
}

\author{Ken-ichi Kubo}

\address{School of Science, Tokyo Metropolitan University,\\
1-1 Minami-osawa, Hachioji, Tokyo 192-0397, Japan  
E-mail:kubo@comp.metro-u.ac.jp }

\author{Katsuhiko Suzuki}
\address{Division of Liberal Arts, Numazu College of Technology,\\
3600 Ooka, Numazu, Shizuoka 410-8501, Japan}
%E-mail:ksuzuki@numazu-ct.ac.jp}  

\maketitle

\abstracts{
We discuss interplay of the scalar and 
vector diquarks which decide the characteristics of the spin polarization as 
functions of $p_T$ and $x_F$ in the wide kinematical range.  We demonstrate a 
significant 
effect of quark masses on the pair-creation probabilities from vacuum in $pp \to \Sigma^- X$ and 
$\gamma p \to \Lambda X$ collisions, and also a remarkable 
correction induced by the 
diquark form factor in $ \Sigma^- p \to \Lambda X $ collision. 
}

\section{Introduction; the global model}

% 
%
%The high energy hadron-hadron collisions produce various hadrons and anti-hadrons 
%through rearrangement processes of various quarks and anti-quarks. 
%We have further 
%variety of the rearrangements when we use other injectiles such as photons, electrons and neutrinos. 
%In other word, we can produce a certain hadron say  through collisions induced by the various injectiles.

The high energy hadron-hadron collisions produce various hyperons, 
in which large spin polarizations perpendicular to a production plane are observed 
at high Feynman-$x$ $(x_F)$ and low transverse momentum $p_T$.   
%Similar polarizations of hyperons are also found in the collisions by 
%other injectiles such as photons, electrons and neutrinos.   
%We have further 
%variety of the collisions when we use other injectiles such as photons, electrons and neutrinos, 
%and found similar polarizations in the final state.  
For these collisions, we have proposed a global model on the basis of quark 
rearrangements concept\cite{1}. 
%The model would be able to apply consistently to the 
%description of 
%spin observables of hadrons produced by the various collisions.  
%Since the measured polarizations are  significant only in the forward (large $x_F$) direction, 
%it is necessary to propose possible mechanism and dynamics which contribute to the 
%forward hadron production 
%and bring about spin asymmetry.  
%To do this, 
In order to produce a final state hyperon at high $x_F$, 
we consider possible rearrangement processes allowed by the SU(6) symmetry, 
where valence quarks (diquarks) from the injectile combine with sea quarks created from 
vacuum.
% so as to 
%which contains maximum numbers of the valence quarks of the injectile and sea quarks 
%created from vacuum.    
%to produce .   
Evaluation of rearrangement amplitudes in terms of relativistic quarks and diquarks with the 
confinement force in fact provides the spin asymmetry.  
We assume that the polarization is principally 
brought about by the rearrangement mechanism,   
while all other processes such as the standard fragmentation, 
which usually comes from showers of partons rather than the valence quark of the injectile and 
thus negligible at high $x_F$, 
are spin-independent.  
If this concept would work for the various hadron productions, we can 
consistently apply it to evaluations of the spin observables by any 
injectile collisions.  
%Then we refer the model as a global model. 

We have applied the model to the various data existing and obtained a general success 
for the spin observables induced by the strong\cite{1,2}, electromagnetic\cite{3,4,5} and 
weak\cite{4,5} interactions, and also for the several so-called puzzling cases\cite{6}.  
%Of course, even this model can work for a wide range of the hadron spin observables, 
%its basic and analytical details should be theoretically studied in future. 
In the presentation, from our recent works, we have concerned about the two indications; 
the quark masses mechanisms 
(QMM) and the form factor dynamics (FFD), for the polarizations in 
$pp \to \Sigma ^- X$, $\gamma p \to \Lambda X$  and  
$\Sigma^- p \to \Lambda X$ collisions.

\vspace{-0.1cm}
\section{The quark masses mechanisms}
%\subsection{The spin polarization of $pp \to \Sigma^- X$ at 400 GeV/c}
$\cdot$\underline{ The spin polarization of $pp \to \Sigma^- X$ at 400 GeV/$c$}

Our global model calculation is compared to the data producing $\Sigma^-$ 
in Fig.1 and it predicts a large negative polarization contrary to 
the positive values in observation.  
In this process, a $d$-valence quark from the injectile $p$ combines with scalar $(ds)^0$
or vector $(ds)^1$ diquarks created from vacuum.   
The discrepancy arises from too much enhancement of the vector diquark  
contribution.  For resolving the problem, we shall take into account the mass dependence of 
probabilities to create the sea diquarks.  
Within our framework, sea quarks are assumed to be produced non-perturbatively by the string 
breaking in the 
hadronization process.  
So far in the previous works\cite{1,3},  we took the same production rate for the creation of the 
all kinds of quarks.  However, it has been suggested that probabilities of the pair-creation 
certainly depend on their masses\cite{8}.  
Indeed, a simple calculation yields a rate of the pair-creation as,
%
%\begin{equation} 
$C^i_{QMM} = \mbox{exp}[-\pi (m^2_i + k_T^2) / \kappa ] \,  $, 
%\label{1}
%\end{equation}
%
where $m$, $k_T$, $\kappa$ are the constituent mass of quark(s) $i$, transverse momentum, and the string 
tension 1GeV$/$fm, respectively.  
%attempt extension of the Schwinger's porposal
%which has been widely used in the calculation of hadron production rates. Its foundation was 
%originated to the fact that the quark and diquark production rates from the vacuum strongly 
%depend on the masses of quarks produced. On the one hand, in our case the different masses appear. 
For the diquark creation we concern, 
the diquark mass  depends on its internal spin state,  and the 
scalar diquark mass $m_S$ is less than the vector one $m_V$.  This is due to a property of 
the interaction acting between the two quarks; attractive for $S=0$, while repulsive for $S=1$.
With $m_S = 0.8$GeV and $m_V = 1 $GeV, 
the spin polarization in our model\cite{1} is then corrected into the form
\begin{equation} 
P(x_F , p_T) = \frac{ R_S \langle C^S_{QMM} \sigma^S_{dep} \rangle  
+ R_V \langle C^V_{QMM} \sigma^V_{dep} \rangle}
{ \langle C^S_{QMM} \sigma^S_{ind} \rangle  
+  \langle C^V_{QMM} \sigma^V_{ind} \rangle}
\end{equation}
The correction is expected to reduce the vector diquark contribution dramatically and in 
fact we have found a significant effect for improvement of the calculation as we 
can see in Fig.1.

\vspace{0.3cm}

%\subsection{The spin polarization of $\gamma p \to \Lambda X$ at $\nu = 16$GeV}
\noindent
$\cdot$\underline{ The spin polarization of $\gamma p \to \Lambda X$ at $\nu = 16$GeV}

Further application of QMM has been made for the  
spin polarization of  $\gamma p \to \Lambda X$ extracted from HERMES data.   
%Using the suitable kinematical cut 
%based on the MC simulation,  one can extract the data of $\Lambda$ polarization produced by the 
%quasi-real $\gamma$ in the forward direction.   
In this case,  $u,d,s$ quarks from $\gamma$ contribute to the  
$\Lambda$ production through the rearrangements $u+(ds)^{0,1}$, $d+(us)^{0,1}$, and $s + (ud)^0$.  
Our original model predicts the dashed curve in Fig.2, which is too big with 
negative-sign contrary to positive-sign in the observation. Then we include QMM 
and got the solid curve which now well reproduces the observation 
with $m^S_{ud} = 0.6$GeV and $m^V_{us}=m^S_{us}=1$GeV.

%In this way, we have found that the quark masses mechanisms also provides a 
%significant effect for the hadron production rates for the photoproduction by 
%reflecting difference in masses of scalar and vector diquarks created from the vacuum.

\vspace{-0.3cm}
\begin{figure}[htb]
\begin{minipage}[t]{54mm}
\epsfxsize=5.cm   
\epsfbox{fig1.eps} 
\caption{Polarization of $pp \to \Sigma^-X$.  The dashed and solid curves show the results without 
and with QMM.}
\end{minipage}
\hspace{\fill}
\begin{minipage}[t]{54mm}
\epsfxsize=5.cm   
\epsfbox{fig2.eps}
\caption{Polarization of $\gamma p \to \Lambda X$.  Notations are the same as those of Fig.1.
}
\end{minipage}
\end{figure}

\vspace{-0.5cm}

\section{The form factor dynamics: $\Sigma^- p \to \Lambda X$ at 340 GeV/$c$}

The full observation has been reported for the polarization of $\Sigma^- p \to \Lambda X$  
in a wide kinematical range.  The most notable point in the observation is change 
of sign of its $p_T$ distribution around $x_F=0.4$, of which shape may not be reproduced 
by any theoretical calculation using the existing models.  Our straight 
forward calculation with the rearrangements $(ds)^{0}+u$ and $(ds)^{1}+u$ 
predicts only a monotonically 
increasing result in Fig.3a due to 
the dominance of the positive $(ds)^1$ contribution.  
%The QMM does not change the calculation.  

Then, we consider here property of the diquark wave functions in the 
configuration space, which should be different between the scalar $(ds)^{0}$  and vector $(ds)^{1}$
diquarks due to attractive or repulsive nature of the color magnetic force.  
%force acting between the 
%two quarks, respectively.   
This interaction may induce a difference of sizes of diquarks, $r_S$ and $r_V$, and 
indeed such a difference is observed within the rigorous three-body calculation of the baryon 
structure\cite{9}.

For taking it into account, it may be reasonable to introduce a form factor for the
quark-diquark interaction vertex.   We take a form $C^{S,V}_{FFD} = \mbox{exp} (-r^2_{S,V} p^2_T)$ 
to be multiplied to each cross section, similarly to eq. (1), with $r_S=0.5$fm and $r_V=0.7$fm.  
%Introduction of FFD enhances the scalar contributions at larger $p_T$.  
We can see in Fig.3b that the FFD correction reduces the monotonical increase of 
the vector component previously seen.   Therefore the new calculation with 
 FFD provides a remarkable change and improves the calculation

\vspace{-0.3cm}
\begin{figure}[htb]
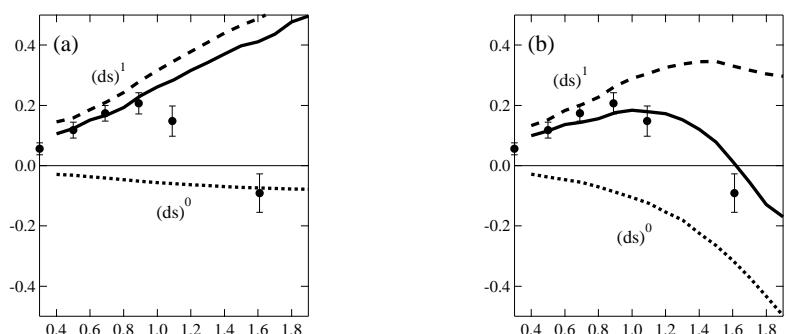

\begin{minipage}[t]{56mm}
\epsfxsize=5.0cm   %width of figure - will enlarge/reduce the figures
\epsfbox{fig3a.eps} 
%\caption{$\Sigma ^- p \to \Lambda X$ without FFD.  }
\end{minipage}
\hspace{\fill}
\begin{minipage}[t]{56mm}
\epsfxsize=5.0cm   %width of figure - will enlarge/reduce the figures
\epsfbox{fig3b.eps}
\end{minipage}
\vspace{-0.8cm}
\caption{Polarization of $\Sigma ^- p \to \Lambda X$ without FFD (left) with FFD (right).}
\end{figure}

\vspace{-0.5cm}

\section{Conclusion}
We discussed the inclusive hyperon productions and the spin polarizations. 
We concentrated on the role of interplay of the scalar and vector diquarks. 
It has been found that, in general, the vector component shows a too much predominance. 
Then we have considered the quark masses mechanisms which provide  
a significant effect to improve the calculations by 
reflecting difference in masses of diquarks created from vacuum.
The form factor dynamics due to the diquark internal structure has been also 
found to provide a remarkable change of the calculation.

\end{document}